\documentclass[preprint,aps]{revtex4-2}

\usepackage{amsmath}
\usepackage{amsfonts}
\usepackage{amssymb}
\usepackage{graphicx}
\usepackage{nicefrac}

\begin{document}


\title{Inverse variational problem for equations in the Riccati chain}


\author{Pranab Sarkar}
\email[]{pranab.sarkar@visva-bharati.ac.in}
\affiliation{Department of Chemistry, Visva-Bharati University, Santiniketan 731235, West Bengal, India}

\author{Pratik Majhi}
\email[]{iampratikmajhi@gmail.com}
\affiliation{Department of Mathematics, Visva-Bharati University, Santiniketan 731235, West Bengal, India}

\author{Madan Mohan Panja}
\email[]{madanpanja2005@yahoo.co.in}
\affiliation{Department of Mathematics, Visva-Bharati University, Santiniketan 731235, West Bengal, India}

\author{Benoy Talukdar}
\email[]{binoy123@bsnl.in}
\affiliation{Department of Physics, Visva-Bharati University, Santiniketan 731235, West Bengal, India}



\begin{abstract}
The nonstandard Lagrangian representations of Ricatti and Riccati-type equations that exist in the literature cannot be obtained using Helmholtz solution of the inverse problem. In this work we consider Riccati and higher-order Riccati equations and construct their  standard Lagrangian representation by using a simple variant of the Helmholtz theory. We make use of the self-adjoint form of the linear equations corresponding to odd-order equations in the Riccati chain to provide a symmetry-based approach for the solution of  inverse  problem. Explicit results presented for Lagrangians of the first and third-order Riccati equations show that one cannot Hamiltonize the Riccati family of equations by the traditional method used in classical mechanics.
\end{abstract}


\maketitle

\section{Introduction}
The $N$th order equation in the Riccati hierarchy can be written as \cite{grundland1999higher}
\begin{equation}
\vartheta^{N}\omega(x)+\sum_{j=1}^{N}\alpha_j(x)\left(\vartheta^{j-1}\omega(x)\right)+\alpha_0(x)=0
\label{psbtrhdrafteq1}
\end{equation}
with the differential operator
\begin{equation}
\vartheta=\frac{d}{dx}+\omega(x).
\label{psbtrhdrafteq2}
\end{equation}
Here $\alpha_j, \, j=0,1,\ldots,N$, stand for $N+1$ well behaved arbitrary functions of $x$. Clearly, for $N=1$, the expression  in \eqref{psbtrhdrafteq1} gives the well known Riccati equation
\begin{equation}
\frac{d\omega(x)}{dx}+\omega(x)^2+\alpha_1(x)\omega(x)+\alpha_0(x)=0
\label{psbtrhdrafteq3}
\end{equation}
while for $N>1$ we get from it higher-order Riccati equations which, together with \eqref{psbtrhdrafteq3}, form the so-called Riccati chain. It is interesting to note that each equation in the Riccati chain  obtained from \eqref{psbtrhdrafteq1} can be linearized by the use of the Cole-Hopf transformation \cite{cole1951quasi,hopf1950partial}
\begin{equation}
\omega(x)=\frac{d\, \ln{y(x)}}{dx}
\label{psbtrhdrafteq4}
\end{equation}
such that use of \eqref{psbtrhdrafteq4} in \eqref{psbtrhdrafteq1} gives the linear equation
\begin{equation}
\sum_{j=0}^{N}\alpha_j(x)\frac{d^j\, y(x)}{dx^j}+\frac{d^{N+1}\, y(x)}{dx^{N+1}}.
\label{psbtrhdrafteq5}
\end{equation}
The object of the present work is  to construct  Lagrangian representations of the  linear equations in  \eqref{psbtrhdrafteq5} and  then  make judicious use of \eqref{psbtrhdrafteq4} to find the Lagrangians for the associated Riccati and/or higher Riccati equation.

In classical mechanics the standard Lagrangian is represented by $L=T-V$, where $T$ is the kinetic energy of the system modeled by the equation and $V$, the corresponding potential function. In addition to the traditional Lagrangian $L$ used in classical mechanics, a new type of Lagrangian functions was proposed \cite{carinena2005lagrangian} for the second-order Riccatti and some dissipative-like autonomous differential equations related to the generalized Riccati equation. These do involve neither $T$ nor $V$. Despite that, the sought Lagrangian functions lead to equations of motion via Euler-Lagrange equations. As a result these Lagrangians were qualified as ‘non-standard’. The non-standard Lagrangians do not have a natural space in the theory of  inverse problem \cite{von1887ueber,saha2014inverse} in the calculus of variation. We are, however, interested here in the inverse variational problem to provide Lagrangian representations of equations in the Ricatti chain. For brevity, we shall refer to the associated Lagrangian as standard one.  The present study is expected to play a role in several areas of physics  since Riccati equation is related to many equations of fundamental  physics \cite{schuch2014relations,schuch2018quantum}. In the recent past Faraoni \cite{faraoni2022helmoltz} reported a solution of the inverse problem for only the Riccati equation by taking recourse to the use of an analogy with the Friedmann  equation \cite{liddle2015introduction}. This method seems to be very special and unlikely to be applicable for other equations in the Riccati chain.

From \eqref{psbtrhdrafteq1} and \eqref{psbtrhdrafteq5} we see that linear equation corresponding  Riccati equation is of order two, while other equations in the Riccati chain are associated with higher-order linear equations. Thus the problem of constructing Lagrangian for higher members in the Riccati chain  should be treated within the framework of a generalized mechanics in which the line integral implied in the definition of action depends on derivatives of generalized coordinates higher than the first. Let $L(x,y(x),y'(x),y''(x),\ldots,y^{(n)}(x))$, with $y^{(n)}(x)=\frac{d^n\, y(x)}{dx^n}$ denote the Lagrangian of a one dimensional system of order $n$. The corresponding Euler-Lagrange equation is given by \cite{cournat1975mimpv1adh}
\begin{equation}
\sum_{i=0}^{n}(-1)^i\frac{d^i}{dx^i}\frac{\partial L}{\partial y^{(i)}}=0.
\label{psbtrhdrafteq6}
\end{equation}
Clearly, the value of $n$ in \eqref{psbtrhdrafteq6} determines the order of the generalized Euler-Lagrange equation. It is well known that the standard Lagrangian $L$ in  \eqref{psbtrhdrafteq6} for $n=1$ leading to second-order differential equation is regular or degenerate if the associated Hessian or functional determinant \cite{santilli2013foundations} is non-null with the possible exception of a finite number of isolated zeros. Otherwise, $L$ is non-degenerate. Similarly, non-zero and zero values of the functional determinant of the higher-order Lagrangian determine if it is degenerate or non-degenerate. In looking for a generalization of the Ostrogradsky theorem, Motohashi and Suyama \cite{motohashi2015third} have clearly pointed out that Lagrangians for even-order linear differential equations are regular while those of odd-order equations are non-degenerate. The odd-order Riccati and/or Riccati type equations, via Cole-Hopf transformation \cite{cole1951quasi,hopf1950partial} lead to even order linear equations. Consequently, Lagrangians of these  equations will be regular. In the following we restrict ourselves to  odd-order Riccati  equations only.

The Lagrangian of any mechanical system governed by a second-order linear differential equation can be found by solving the so-called Helmholtz problem \cite{von1887ueber}. It is, however, not straightforward to adapt the approach to deal with higher-order systems. It appears that investigation in the inverse variational problem for such systems has yet remained largely unexplored \cite{subrate2010inequivalentphdthesislatpi}. In view of this, we shall make use of certain symmetry considerations \cite{hojman1984symmetries} to provide a simple recipe  to compute Lagrangians of higher-order systems. We postulate that the Lagrangian of a higher-order system is proportional to the self adjoint form of its equation of motion. This simple minded solution of the inverse problem sought  by us has an old root in the scientific literature \cite{hojman1983shortcut}.

The theory of self-adjoint  differential equation for the second-order equation  can be found in any standard text book on mathematical physics \cite{arfken1972mathematicalhjwmmfpse}. This is, however, not true for higher-order equations. We thus begin Sec.~\ref{psbtrhdraftlabelsec2}  with a  brief introduction to algebraic properties of higher-order self-adjoint systems  and  then present an ansantz  to solve the associated inverse vatiational problem. In Sec.~\ref{psbtrhdraftlabelsec3} we provide some useful checks for the validity of the proposed ansatz in terms of two well-known physically important examples and then demonstrate how our approach could be judiciously employed to find a standard Lagrangian representation of the Riccati equation.  Sec.~\ref{psbtrhdraftlabelsec4} is devoted  to deal with similar  problems for  higher-order equations  in the Riccati chain. Here we pay special  attention to the third-order Riccati equation with an associated fourth-order linear equation and present explicit result for its Lagrangian function. In Appendix  we provide a general expression for the Lagrangian of $(2n-1)$th order equation, $n\geq1$, in the Riccati chain. Finally, in Sec.~\ref{psbtrhdraftlabelsec5}, we summarize our outlook on the present work and make some concluding remarks.
\section{Self-adjoint differential equations and inverse variational problem \label{psbtrhdraftlabelsec2}}

Let $M(y)$ stand for the differential expression 
\begin{equation}
M(y)=r_0(x)\frac{d^ny(x)}{dx^n}+r_1(x)\frac{d^{n-1}y(x)}{dx^{n-1}}+\cdots+r_n(x)y(x),
\label{psbtrhdrafteq7}
\end{equation}
where $n$ is any positive integer. The coefficients $r$ are real or complex function of the real variable $x$, defined in an interval $a\leq x\leq b$. The relation $M(y)=0$ defines an $n$th order non self-adjoint linear differential equation. The general adjoint system associated with the differential operator  in \eqref{psbtrhdrafteq7} is written as
\begin{equation}
N(z)=(-1)^n\frac{d^n(r_0(x)z(x))}{dx^n}+(-1)^{n-1}\frac{d^{n-1}(r_1(x)z(x))}{dx^{n-1}}+\cdots+r_n(x)z(x).
\label{psbtrhdrafteq8}
\end{equation}
The adjoint operator $N(z)$ is connected with $M(y)$ by the Lagrangian identity
\begin{equation}
zM(y)-yN(z)=\frac{dp(y,z)}{dx},
\label{psbtrhdrafteq9}
\end{equation}
where $p(y,z)$ stands for a bilinear form in  $y(x),y'(x),\ldots,y^{(n-1)}(x);z(x),z'(x),\ldots,z^{(n-1)}(x)$ with coefficients which are functions of  $x$. Ince \cite{ince1956ordinary} has carefully demonstrated how to exploit \eqref{psbtrhdrafteq7}, \eqref{psbtrhdrafteq8} and \eqref{psbtrhdrafteq9} to construct the self-adjoint equation for a given higher-order linear ordinary differential equation. We shall now present a recipe for the solution of the inverse variational problem for a higher-order equation, where its self-adjoint form plays a crucial role.

The relation between symmetries of a Lagrangian and conserved quantities of the corresponding equation of motion is a very well known result in classical mechanics. But it is less well known that the symmetries of the equation of motion form a larger set than that of the Lagrangian. Hojman \cite{hojman1984symmetries} supplemented the well-known concept of  s-equivalence \cite{currie1966q} by a new kind of symmetry such that the resulting set of Lagragian symmetries coincides with the symmetries of the equations of motions. In this context it was established that a second-order Lagrangian is proportional to a second-order differential equation which follows from the traditional first-order Lagrangian. The constant of proportionality is determined by a gauge term that reduces the expression for the second-order Lagrangian to the first-order one. We postulate that the $n$th order Lagrangian bears a similar relation with the $n$th differential equation  such that the ansatz
\begin{equation}
L\left(x,y(x),y'(x),y''(x),\ldots,y^{(n)}(x)\right)=y(x) M_s(y),
\label{psbtrhdrafteq10}
\end{equation} 
provides a simple solution  for the inverse variational  problem for higher-order systems. In equation \eqref{psbtrhdrafteq10}, $M_s(y)$ stands for the self-adjoint  differential expression corresponding to $M(y)$.
\section{Lagrangian function for the Riccati equation \label{psbtrhdraftlabelsec3}}
Before  we employ \eqref{psbtrhdrafteq10} to compute the   Lagrangian  function of the Riccati  equation, we would  like to present some useful checks on  its  validity  in terms of simple case studies. To that let us consider the following second- and fourth-order self-adjoint  equations.
\begin{equation}
y''(x)+y(x)=0
\label{psbtrhdrafteq11}
\end{equation}
and
\begin{equation}
y^{(4)}(x)+y(x)=0.
\label{psbtrhdrafteq12}
\end{equation}
Because of \eqref{psbtrhdrafteq10} the Lagrangians for \eqref{psbtrhdrafteq11} and \eqref{psbtrhdrafteq12} are given by
\begin{equation}
L\left(x,y(x),y'(x),y''(x)\right)=y(x)\left(y''(x)+y(x)\right)
\label{psbtrhdrafteq13}
\end{equation}
and
\begin{equation}
L\left(x,y(x),y'(x),y''(x),y'''(x),y^{(4)}(x)\right)=y(x)\left(y^{(4)}(x)+y(x)\right).
\label{psbtrhdrafteq14}
\end{equation}
When a gauge term $\frac{d}{dx}\left(y(x)y'(x)\right)$ is subtracted from the second-order Lagrangian in \eqref{psbtrhdrafteq13} we get the familiar first-order Lagrangian for \eqref{psbtrhdrafteq11}. The gauge term  for a  fourth-order Lagrangian is $\frac{d}{dx}\left(y(x)y'''(x)\right)$ \cite{caratheodory1982calculus} which when subtracted   from \eqref{psbtrhdrafteq14} leads to a third-order Lagrangian. It is easy to verify that the constructed lower order Lagrangian via the appropriate Euler-Lagrange equation reproduces the fourth-order equation in \eqref{psbtrhdrafteq12}. These results suggest that application of our solution of the inverse problem to Riccati equation is now in order.

From \eqref{psbtrhdrafteq5} the linear equation corresponding to the Riccati equation \eqref{psbtrhdrafteq3} is obtained as
\begin{equation}
y''(x)+\alpha_1(x)y'(x)+\alpha_0(x)y(x)=0.
\label{psbtrhdrafteq15}
\end{equation}
Equation \eqref{psbtrhdrafteq15} is not self adjoint. But it can be reduced to  the self-adjoint form by multiplying with $e^{\int \alpha_1(x)dx}$ \cite{cournat1975mimpv1adh} so as to write the Lagrangian for \eqref{psbtrhdrafteq15} as 
\begin{equation}
L=y(x)e^{\int \alpha_1(x)dx}\left(y''(x)+\alpha_1(x)y'(x)+\alpha_0(x)y(x)\right).
\label{psbtrhdrafteq16}
\end{equation}
From the Cole-Hopf transformation \eqref{psbtrhdrafteq4} we write
\begin{equation}
y(x)=a e^{\int \omega(x)dx},
\label{psbtrhdrafteq17}
\end{equation}
where $a$ is a constant of integration, Equations \eqref{psbtrhdrafteq16} and \eqref{psbtrhdrafteq17} give the Lagrangian for the Riccati equation.
\begin{equation}
L=a^2 e^{\int \alpha_1(x)dx+2\int \omega(x) dx} \left(\omega'(x)+\omega^2(x)+\alpha_1(x) \omega(x)+\alpha_0(x)\right).
\label{psbtrhdrafteq18}
\end{equation}
It is clear from \eqref{psbtrhdrafteq18} that as with the Lagrangian of the linear equation \eqref{psbtrhdrafteq15}, the Lagrangian for the Riccati equation, which is essentially nonlinear, also involves its own equation of motion. It is easy to verify that $L$ in \eqref{psbtrhdrafteq18} leads to the Riccati equation via  the traditional first-order Euler-Lagrange equation. We shall now extend the above approach to higher-order equations in the Riccati chain.
\section{Lagrangian functions for odd-order equations in the Riccati Chain \label{psbtrhdraftlabelsec4}}

Next to Riccati equation we are interested  in the third-order equation in the Riccati chain presumably because the linear equation corresponding to it  is of even order. Using $N=3$ in \eqref{psbtrhdrafteq1} and \eqref{psbtrhdrafteq5} we get results for the  Riccati-3 and corresponding linear equation in the following form.
\begin{eqnarray}
\frac{d^3\omega}{dx^3}+\left(\alpha_3(x)+4\omega(x)\right)\frac{d^2\omega}{dx^2}+3\left(\frac{d\omega}{dx}\right)^2+\left(6\omega(x)^2+3\alpha_3(x)\omega(x)+\alpha_2(x)\right)\frac{d\omega}{dx}+\omega(x)^4\nonumber \\+\alpha_3(x)\omega(x)^3+\alpha_2(x)\omega(x)^2+\alpha_1(x)\omega(x)+\alpha_0(x)=0 \hspace{2cm}
\label{psbtrhdrafteq19}
\end{eqnarray}
and
\begin{equation}
\frac{d^4y}{dx^4}+\alpha_3(x) \frac{d^3y}{dx^3}+\alpha_2(x)\frac{d^2y}{dx^2}+\alpha_1(x)\frac{dy}{dx}+\alpha_0(x)y=0.
\label{psbtrhdrafteq20}
\end{equation}
Clearly, Eq.~\eqref{psbtrhdrafteq20} is a special form of the more general equation
\begin{equation}
r_0(x)\frac{d^4y}{dx^4}+r_1(x) \frac{d^3y}{dx^3}+r_2(x)\frac{d^2y}{dx^2}+r_3(x)\frac{dy}{dx}+r_4(x)y=0.
\label{psbtrhdrafteq21}
\end{equation}
obtained from $M(y)=0$ for $n=4$. The self-adjoint form of \eqref{psbtrhdrafteq21} can be written as \cite{beccar2019analytic}
\begin{equation}
\left(r_0(x)y''(x)\right)''+\left(q(x)y'(x)\right)'+r_4(x)y(x)=0
\label{psbtrhdrafteq22}
\end{equation}
 with the following conditions
 \begin{equation}
 r_1(x)=2r'_{0}(x), \, q(x)=r_2(x)-r''_{0}(x)\, \, \, \textrm{and} \, \, q'(x)=r_3(x).
 \label{psbtrhdrafteq23}
 \end{equation}
 In view of \eqref{psbtrhdrafteq10}  we use \eqref{psbtrhdrafteq22} to write  the Lagrangian  of the self-adloint equation in the  form 
 \begin{equation}
 L=y(x)\left(r_0(x)y^{(4)}(x)+2r'_0(x)y^{(3)}(x)+r''_0(x)y''(x)+q(x)y''(x)+q'(x)y'(x)+r_4(x)y(x)\right).
  \label{psbtrhdrafteq24}
 \end{equation}
 It is straightforward to verify that $L$ in \eqref{psbtrhdrafteq24} via the fourth-order Euler-Lagrange equation obtained from 
\eqref{psbtrhdrafteq6} reproduces \eqref{psbtrhdrafteq22}. Making use of \eqref{psbtrhdrafteq23} we rewrite \eqref{psbtrhdrafteq24}  to read 
\begin{equation}
L=y(x)\left(r_0(x)y^{(4)}(x)+r_1(x)y^{(3)}(x)+r_2(x)y''(x)+r_3(x)y'(x)+r_4(x)y(x)\right).
\label{psbtrhdrafteq25}
\end{equation}
Comparing \eqref{psbtrhdrafteq20} and \eqref{psbtrhdrafteq21} we have $r_0(x)=1, \, r_1(x)=\alpha_3(x), \, r_2(x)=\alpha_2(x),\, r_3(x)=\alpha_1(x)$ and $r_4(x)=\alpha_0(x)$. When these relations are substituted in \eqref{psbtrhdrafteq25} we get the Lagrangian for the fourth-order linear equation corresponding third-order equation in the Riccati chain.
\begin{equation}
L=y(x)\left(y^{(4)}(x)+\alpha_3(x) y^{(3)}(x)+\alpha_2(x)y''(x)+\alpha_1(x)y'(x)+\alpha_0(x)y(x)\right).
\label{psbtrhdrafteq26}
\end{equation}
We can now combine \eqref{psbtrhdrafteq17} and \eqref{psbtrhdrafteq26} to obtain the Lagrangian of the third-order Riccati equation as
\begin{equation}
L=a^2e^{2\int \omega(x)dx}\left(S_1(x)+S_2(x)\right)
\label{psbtrhdrafteq27}
\end{equation}
with
\begin{equation}
S_1(x)=\omega^{(3)}(x)+4\omega(x)\omega''(x)+\alpha_3(x)\omega''(x)+3\omega'(x)^2+6\omega(x)^2\omega'(x)+3\alpha_3(x)\omega(x)\omega'(x)
\label{psbtrhdrafteq28}
\end{equation}
and
\begin{equation}
S_2(x)=\alpha_2(x)\omega'(x)+\omega(x)^4+\alpha_3(x)\omega(x)^3+\alpha_2(x)\omega(x)^2+\alpha_1(x)\omega(x)+\alpha_0(x).
\label{psbtrhdrafteq29}
\end{equation}
It is straightforward to verify that the result in \eqref{psbtrhdrafteq29} follows from \eqref{psbtrhdraftappendixeq8} in the Appendix for $n=2$.
\section{Concluding remarks \label{psbtrhdraftlabelsec5}}
The Newtonian approach to classical mechanics is a force-momentum formulation with cause and effect embedded in it. On the other hand, the Lagrangian approach is developed in terms of kinetic and potential energies and makes use of a single scalar quantity called the Lagragion. In the recent past, connections were established between the Newton’s laws of motion and Riccati equation \cite{nowakowski2002newton}. Here we look for Lagrangian representation of equations in the Riccati chain.  The explicit result presented for the Riccati equation exhibits that it follows from a first-order Lagrangian depending  linearly on $\omega'(x)$ rather than quadratically. This provides an awkward analytic constraint to Hamiltonize the Riccati system because the quantity $\frac{dL}{d\omega'(x)}$ cannot be inverted for $\omega'(x)$. Consequently, we need to take recourse to the use of the Dirac’s theory of constraints \cite{dirac1964loqmlgsmsn2} to obtain the necessary Hamiltonian formulation. We have an identicaql situation for the third-order Riccati equation because the Lagrangian in \eqref{psbtrhdrafteq27} is also linear in $\omega^{(3)}$. This is. In general, true for all odd-order equations in the Riccati chain. It will be quite interesting to look for Hamiltonian formulation of Riccati equations because besides classical mechanics Hamiltonian formulation plays important roles in statistical physics and quantum mechanics \cite{sudarshan1974classical}. Finally, we conclude by noting that in the present study we concentrated on the Lagrangian representation for half of the equations in the Ricatti chain. As regards the other half, we noted that each even-order equation in the Riccati chain leads to of odd order linear equation. This did not permit us to look for solution of the inverse problem for  even-order Riccati-type equations. But we believe that the work of Motohashi and Suyama \cite{motohashi2015third} may be of some help in respect of this.

\bibliography{ivpref}

\providecommand{\noopsort}[1]{}\providecommand{\singleletter}[1]{#1}%
\begin{thebibliography}{24}%
\makeatletter
\providecommand \@ifxundefined [1]{%
 \@ifx{#1\undefined}
}%
\providecommand \@ifnum [1]{%
 \ifnum #1\expandafter \@firstoftwo
 \else \expandafter \@secondoftwo
 \fi
}%
\providecommand \@ifx [1]{%
 \ifx #1\expandafter \@firstoftwo
 \else \expandafter \@secondoftwo
 \fi
}%
\providecommand \natexlab [1]{#1}%
\providecommand \enquote  [1]{``#1''}%
\providecommand \bibnamefont  [1]{#1}%
\providecommand \bibfnamefont [1]{#1}%
\providecommand \citenamefont [1]{#1}%
\providecommand \href@noop [0]{\@secondoftwo}%
\providecommand \href [0]{\begingroup \@sanitize@url \@href}%
\providecommand \@href[1]{\@@startlink{#1}\@@href}%
\providecommand \@@href[1]{\endgroup#1\@@endlink}%
\providecommand \@sanitize@url [0]{\catcode `\\12\catcode `\$12\catcode
  `\&12\catcode `\#12\catcode `\^12\catcode `\_12\catcode `\%12\relax}%
\providecommand \@@startlink[1]{}%
\providecommand \@@endlink[0]{}%
\providecommand \url  [0]{\begingroup\@sanitize@url \@url }%
\providecommand \@url [1]{\endgroup\@href {#1}{\urlprefix }}%
\providecommand \urlprefix  [0]{URL }%
\providecommand \Eprint [0]{\href }%
\providecommand \doibase [0]{https://doi.org/}%
\providecommand \selectlanguage [0]{\@gobble}%
\providecommand \bibinfo  [0]{\@secondoftwo}%
\providecommand \bibfield  [0]{\@secondoftwo}%
\providecommand \translation [1]{[#1]}%
\providecommand \BibitemOpen [0]{}%
\providecommand \bibitemStop [0]{}%
\providecommand \bibitemNoStop [0]{.\EOS\space}%
\providecommand \EOS [0]{\spacefactor3000\relax}%
\providecommand \BibitemShut  [1]{\csname bibitem#1\endcsname}%
\let\auto@bib@innerbib\@empty
\bibitem [{\citenamefont {Grundland}\ and\ \citenamefont
  {Levi}(1999)}]{grundland1999higher}%
  \BibitemOpen
  \bibfield  {author} {\bibinfo {author} {\bibfnamefont {A.~M.}\ \bibnamefont
  {Grundland}}\ and\ \bibinfo {author} {\bibfnamefont {D.}~\bibnamefont
  {Levi}},\ }\bibfield  {title} {\bibinfo {title} {Higher-order riccati
  equations as b{\"a}cklund transformations},\ }\href@noop {} {\bibfield
  {journal} {\bibinfo  {journal} {J. Phys. A: Math-Gen}\ }\textbf {\bibinfo
  {volume} {32}},\ \bibinfo {pages} {3931} (\bibinfo {year}
  {1999})}\BibitemShut {NoStop}%
\bibitem [{\citenamefont {Cole}(1951)}]{cole1951quasi}%
  \BibitemOpen
  \bibfield  {author} {\bibinfo {author} {\bibfnamefont {J.~D.}\ \bibnamefont
  {Cole}},\ }\bibfield  {title} {\bibinfo {title} {On a quasi-linear parabolic
  equation occurring in aerodynamics},\ }\href@noop {} {\bibfield  {journal}
  {\bibinfo  {journal} {Quart. Appl. Math.}\ }\textbf {\bibinfo {volume} {9}},\
  \bibinfo {pages} {225} (\bibinfo {year} {1951})}\BibitemShut {NoStop}%
\bibitem [{\citenamefont {Hopf}(1950)}]{hopf1950partial}%
  \BibitemOpen
  \bibfield  {author} {\bibinfo {author} {\bibfnamefont {E.}~\bibnamefont
  {Hopf}},\ }\bibfield  {title} {\bibinfo {title} {The partial differential
  equation {$u_t+ uu_x= u_{xx}$}},\ }\href@noop {} {\bibfield  {journal}
  {\bibinfo  {journal} {Comm. Pure Appl. Math.}\ }\textbf {\bibinfo {volume}
  {3}},\ \bibinfo {pages} {201} (\bibinfo {year} {1950})}\BibitemShut {NoStop}%
\bibitem [{\citenamefont {Cari{\~n}ena}\ \emph {et~al.}(2005)\citenamefont
  {Cari{\~n}ena}, \citenamefont {Ra{\~n}ada},\ and\ \citenamefont
  {Santander}}]{carinena2005lagrangian}%
  \BibitemOpen
  \bibfield  {author} {\bibinfo {author} {\bibfnamefont {J.~F.}\ \bibnamefont
  {Cari{\~n}ena}}, \bibinfo {author} {\bibfnamefont {M.~F.}\ \bibnamefont
  {Ra{\~n}ada}},\ and\ \bibinfo {author} {\bibfnamefont {M.}~\bibnamefont
  {Santander}},\ }\bibfield  {title} {\bibinfo {title} {Lagrangian formalism
  for nonlinear second-order riccati systems: one-dimensional integrability and
  two-dimensional superintegrability},\ }\href@noop {} {\bibfield  {journal}
  {\bibinfo  {journal} {J. Math. Phy.}\ }\textbf {\bibinfo {volume} {46}},\
  \bibinfo {pages} {062703} (\bibinfo {year} {2005})}\BibitemShut {NoStop}%
\bibitem [{\citenamefont {Helmholtz}(1887)}]{von1887ueber}%
  \BibitemOpen
  \bibfield  {author} {\bibinfo {author} {\bibfnamefont {H.}~\bibnamefont
  {Helmholtz}},\ }\bibfield  {title} {\bibinfo {title} {Ueber die physikalische
  bedeutung des prinicips der kleinsten wirkung},\ }\href@noop {} {\bibfield
  {journal} {\bibinfo  {journal} {J. Reine Angew. Math.}\ }\textbf {\bibinfo
  {volume} {100}},\ \bibinfo {pages} {213} (\bibinfo {year}
  {1887})}\BibitemShut {NoStop}%
\bibitem [{\citenamefont {Saha}\ and\ \citenamefont
  {Talukdar}(2014)}]{saha2014inverse}%
  \BibitemOpen
  \bibfield  {author} {\bibinfo {author} {\bibfnamefont {A.}~\bibnamefont
  {Saha}}\ and\ \bibinfo {author} {\bibfnamefont {B.}~\bibnamefont
  {Talukdar}},\ }\bibfield  {title} {\bibinfo {title} {Inverse variational
  problem for nonstandard lagrangians},\ }\href@noop {} {\bibfield  {journal}
  {\bibinfo  {journal} {Rep. Math. Phys.}\ }\textbf {\bibinfo {volume} {73}},\
  \bibinfo {pages} {299} (\bibinfo {year} {2014})}\BibitemShut {NoStop}%
\bibitem [{\citenamefont {Schuch}(2014)}]{schuch2014relations}%
  \BibitemOpen
  \bibfield  {author} {\bibinfo {author} {\bibfnamefont {D.}~\bibnamefont
  {Schuch}},\ }\bibfield  {title} {\bibinfo {title} {Relations between
  nonlinear riccati equations and other equations in fundamental physics},\
  }\bibfield  {booktitle} {\emph {\bibinfo {booktitle} {J. Phys. Conf. Ser.}},\
  }\href@noop {} {\ \textbf {\bibinfo {volume} {538}},\ \bibinfo {pages}
  {012019} (\bibinfo {year} {2014})}\BibitemShut {NoStop}%
\bibitem [{\citenamefont {Schuch}(2018)}]{schuch2018quantum}%
  \BibitemOpen
  \bibfield  {author} {\bibinfo {author} {\bibfnamefont {D.}~\bibnamefont
  {Schuch}},\ }\href@noop {} {\emph {\bibinfo {title} {Quantum theory from a
  nonlinear perspective}}}\ (\bibinfo  {publisher} {Springer},\ \bibinfo
  {address} {New York},\ \bibinfo {year} {2018})\BibitemShut {NoStop}%
\bibitem [{\citenamefont {Faraoni}(2022)}]{faraoni2022helmoltz}%
  \BibitemOpen
  \bibfield  {author} {\bibinfo {author} {\bibfnamefont {V.}~\bibnamefont
  {Faraoni}},\ }\bibfield  {title} {\bibinfo {title} {Helmoltz problem for the
  riccati equation from an analogous friedmann equation},\ }\href@noop {}
  {\bibfield  {journal} {\bibinfo  {journal} {Eur. Phys. J. C}\ }\textbf
  {\bibinfo {volume} {82}},\ \bibinfo {pages} {13} (\bibinfo {year}
  {2022})}\BibitemShut {NoStop}%
\bibitem [{\citenamefont {Liddle}(2015)}]{liddle2015introduction}%
  \BibitemOpen
  \bibfield  {author} {\bibinfo {author} {\bibfnamefont {A.}~\bibnamefont
  {Liddle}},\ }\href@noop {} {\emph {\bibinfo {title} {An introduction to
  modern cosmology}}}\ (\bibinfo  {publisher} {Wiley},\ \bibinfo {address} {New
  York},\ \bibinfo {year} {2015})\BibitemShut {NoStop}%
\bibitem [{\citenamefont {Courant}\ and\ \citenamefont
  {Hilbert}(1975)}]{cournat1975mimpv1adh}%
  \BibitemOpen
  \bibfield  {author} {\bibinfo {author} {\bibfnamefont {R.}~\bibnamefont
  {Courant}}\ and\ \bibinfo {author} {\bibfnamefont {D.}~\bibnamefont
  {Hilbert}},\ }\href@noop {} {\emph {\bibinfo {title} {Methods in Mathematical
  Physics, Vol. 1}}}\ (\bibinfo  {publisher} {Wiley Eastern Pvt. Ltd.},\
  \bibinfo {address} {New Delhi},\ \bibinfo {year} {1975})\BibitemShut
  {NoStop}%
\bibitem [{\citenamefont {Santilli}(1978)}]{santilli2013foundations}%
  \BibitemOpen
  \bibfield  {author} {\bibinfo {author} {\bibfnamefont {R.~M.}\ \bibnamefont
  {Santilli}},\ }\href@noop {} {\emph {\bibinfo {title} {Foundations of
  Theoretical Mechanics I: the inverse problem in Newtonian mechanics, Vol.
  1}}}\ (\bibinfo  {publisher} {Springer Verlag},\ \bibinfo {address} {New
  York},\ \bibinfo {year} {1978})\BibitemShut {NoStop}%
\bibitem [{\citenamefont {Motohashi}\ and\ \citenamefont
  {Suyama}(2015)}]{motohashi2015third}%
  \BibitemOpen
  \bibfield  {author} {\bibinfo {author} {\bibfnamefont {H.}~\bibnamefont
  {Motohashi}}\ and\ \bibinfo {author} {\bibfnamefont {T.}~\bibnamefont
  {Suyama}},\ }\bibfield  {title} {\bibinfo {title} {Third order equations of
  motion and the ostrogradsky instability},\ }\href@noop {} {\bibfield
  {journal} {\bibinfo  {journal} {Phys. Rev. D}\ }\textbf {\bibinfo {volume}
  {91}},\ \bibinfo {pages} {085009} (\bibinfo {year} {2015})}\BibitemShut
  {NoStop}%
\bibitem [{\citenamefont
  {Ghosh}(2010)}]{subrate2010inequivalentphdthesislatpi}%
  \BibitemOpen
  \bibfield  {author} {\bibinfo {author} {\bibfnamefont {S.}~\bibnamefont
  {Ghosh}},\ }\emph {\bibinfo {title} {Inequivalent Lagrangians and their
  Physical Implications}},\ \href@noop {} {\bibinfo {type} {{Ph.D.} thesis}},\
  \bibinfo  {school} {Visva-Bharati University}, \bibinfo {address}
  {Santiniketan 731235, West Bengal, India} (\bibinfo {year}
  {2010})\BibitemShut {NoStop}%
\bibitem [{\citenamefont {Hojman}(1984)}]{hojman1984symmetries}%
  \BibitemOpen
  \bibfield  {author} {\bibinfo {author} {\bibfnamefont {S.}~\bibnamefont
  {Hojman}},\ }\bibfield  {title} {\bibinfo {title} {Symmetries of lagrangians
  and of their equations of motion},\ }\href@noop {} {\bibfield  {journal}
  {\bibinfo  {journal} {J. Phys. A : Math. Gen.}\ }\textbf {\bibinfo {volume}
  {17}},\ \bibinfo {pages} {2399} (\bibinfo {year} {1984})}\BibitemShut
  {NoStop}%
\bibitem [{\citenamefont {Hojman}\ \emph {et~al.}(1983)\citenamefont {Hojman},
  \citenamefont {Hojman},\ and\ \citenamefont
  {Sheinbaum}}]{hojman1983shortcut}%
  \BibitemOpen
  \bibfield  {author} {\bibinfo {author} {\bibfnamefont {R.}~\bibnamefont
  {Hojman}}, \bibinfo {author} {\bibfnamefont {S.}~\bibnamefont {Hojman}},\
  and\ \bibinfo {author} {\bibfnamefont {J.}~\bibnamefont {Sheinbaum}},\
  }\bibfield  {title} {\bibinfo {title} {Shortcut for constructing any
  lagrangian from its equations of motion},\ }\href@noop {} {\bibfield
  {journal} {\bibinfo  {journal} {Phys. Rev. D}\ }\textbf {\bibinfo {volume}
  {28}},\ \bibinfo {pages} {1333} (\bibinfo {year} {1983})}\BibitemShut
  {NoStop}%
\bibitem [{\citenamefont {Arfken}\ and\ \citenamefont
  {Weber}(2012)}]{arfken1972mathematicalhjwmmfpse}%
  \BibitemOpen
  \bibfield  {author} {\bibinfo {author} {\bibfnamefont {G.~B.}\ \bibnamefont
  {Arfken}}\ and\ \bibinfo {author} {\bibfnamefont {H.~J.}\ \bibnamefont
  {Weber}},\ }\href@noop {} {\emph {\bibinfo {title} {Mathematical methods for
  physicists, Sixth ed.}}}\ (\bibinfo  {publisher} {Elsevier Academic Press},\
  \bibinfo {address} {New York},\ \bibinfo {year} {2012})\BibitemShut {NoStop}%
\bibitem [{\citenamefont {Ince}(1958)}]{ince1956ordinary}%
  \BibitemOpen
  \bibfield  {author} {\bibinfo {author} {\bibfnamefont {E.~L.}\ \bibnamefont
  {Ince}},\ }\href@noop {} {\emph {\bibinfo {title} {Ordinary differential
  equations}}}\ (\bibinfo  {publisher} {Dover Publication},\ \bibinfo {address}
  {New York},\ \bibinfo {year} {1958})\BibitemShut {NoStop}%
\bibitem [{\citenamefont {Currie}\ and\ \citenamefont
  {Saletan}(1966)}]{currie1966q}%
  \BibitemOpen
  \bibfield  {author} {\bibinfo {author} {\bibfnamefont {D.~G.}\ \bibnamefont
  {Currie}}\ and\ \bibinfo {author} {\bibfnamefont {E.~J.}\ \bibnamefont
  {Saletan}},\ }\bibfield  {title} {\bibinfo {title} {q-equivalent particle
  hamiltonians. i. the classical one-dimensional case},\ }\href@noop {}
  {\bibfield  {journal} {\bibinfo  {journal} {J. Math. Phys.}\ }\textbf
  {\bibinfo {volume} {7}},\ \bibinfo {pages} {967} (\bibinfo {year}
  {1966})}\BibitemShut {NoStop}%
\bibitem [{\citenamefont {Carath{\'e}odory}(1967)}]{caratheodory1982calculus}%
  \BibitemOpen
  \bibfield  {author} {\bibinfo {author} {\bibfnamefont {C.}~\bibnamefont
  {Carath{\'e}odory}},\ }\href@noop {} {\emph {\bibinfo {title} {Calculus of
  variations and partial differential equations of the first order, Vol. 2}}}\
  (\bibinfo  {publisher} {Holden-Day Inc.},\ \bibinfo {address} {San
  Fransisco},\ \bibinfo {year} {1967})\BibitemShut {NoStop}%
\bibitem [{\citenamefont {Beccar-Varela}\ \emph {et~al.}(2019)\citenamefont
  {Beccar-Varela}, \citenamefont {Bhuiyan}, \citenamefont {Mariani},\ and\
  \citenamefont {Tweneboah}}]{beccar2019analytic}%
  \BibitemOpen
  \bibfield  {author} {\bibinfo {author} {\bibfnamefont {M.~P.}\ \bibnamefont
  {Beccar-Varela}}, \bibinfo {author} {\bibfnamefont {M.~A.~M.}\ \bibnamefont
  {Bhuiyan}}, \bibinfo {author} {\bibfnamefont {M.~C.}\ \bibnamefont
  {Mariani}},\ and\ \bibinfo {author} {\bibfnamefont {O.~K.}\ \bibnamefont
  {Tweneboah}},\ }\bibfield  {title} {\bibinfo {title} {Analytic methods for
  solving higher order ordinary differential equations},\ }\href@noop {}
  {\bibfield  {journal} {\bibinfo  {journal} {Mathematics}\ }\textbf {\bibinfo
  {volume} {7}},\ \bibinfo {pages} {826} (\bibinfo {year} {2019})}\BibitemShut
  {NoStop}%
\bibitem [{\citenamefont {Nowakowski}\ and\ \citenamefont
  {Rosu}(2002)}]{nowakowski2002newton}%
  \BibitemOpen
  \bibfield  {author} {\bibinfo {author} {\bibfnamefont {M.}~\bibnamefont
  {Nowakowski}}\ and\ \bibinfo {author} {\bibfnamefont {H.~C.}\ \bibnamefont
  {Rosu}},\ }\bibfield  {title} {\bibinfo {title} {Newton’s laws of motion in
  the form of a riccati equation},\ }\href@noop {} {\bibfield  {journal}
  {\bibinfo  {journal} {Phys. Rev. E}\ }\textbf {\bibinfo {volume} {65}},\
  \bibinfo {pages} {047602} (\bibinfo {year} {2002})}\BibitemShut {NoStop}%
\bibitem [{\citenamefont {Dirac}(1964)}]{dirac1964loqmlgsmsn2}%
  \BibitemOpen
  \bibfield  {author} {\bibinfo {author} {\bibfnamefont {P.~A.~M.}\
  \bibnamefont {Dirac}},\ }\href@noop {} {\emph {\bibinfo {title} {Lectures on
  Quantum Mechanics}}},\ Belfer Graduate School Monograph Series No. 2\
  (\bibinfo  {publisher} {Yeshiva University},\ \bibinfo {address} {NY},\
  \bibinfo {year} {1964})\BibitemShut {NoStop}%
\bibitem [{\citenamefont {Sudarshan}\ and\ \citenamefont
  {Mukunda}(1974)}]{sudarshan1974classical}%
  \BibitemOpen
  \bibfield  {author} {\bibinfo {author} {\bibfnamefont {E.~C.~G.}\
  \bibnamefont {Sudarshan}}\ and\ \bibinfo {author} {\bibfnamefont
  {N.}~\bibnamefont {Mukunda}},\ }\href@noop {} {\emph {\bibinfo {title}
  {Classical dynamics: a modern perspective}}}\ (\bibinfo  {publisher}
  {John-Wiley \& Sons, Inc.},\ \bibinfo {address} {NY},\ \bibinfo {year}
  {1974})\BibitemShut {NoStop}%
\end{thebibliography}%

\appendix*
\section{General  expression  for the Lagrangian  of any odd-order equation in the Riccati Chain \label{appendix1} }
From Eq.~\eqref{psbtrhdrafteq1} we write
\begin{equation}
\vartheta^{2n-1}\omega(x)+\sum_{j=1}^{2n-1}\alpha_j(x)\left(\vartheta^{j-1}\omega(x)\right)+\alpha_0(x)=0
\label{psbtrhdraftappendixeq1}
\end{equation}
for the  odd-order Riccati  equations. Here $n\geq 1$. Due to the transformation  in Eq.~\eqref{psbtrhdrafteq4}  the even-order linear equations corresponding to \eqref{psbtrhdraftappendixeq1} are  given by 
\begin{equation}
\frac{d^{2n}y}{dx^{2n}}+\sum_{j=0}^{2n-1}\alpha_j(x) \frac{d^jy}{dx^j}=0.
\label{psbtrhdraftappendixeq2}
\end{equation} 
The general  linear homogeneous ordinary differential equation
\begin{equation}
\sum_{i=0}^{2n}r_{i}(x)\frac{d^{2n-i}y}{dx^{2n-i}}=0
\label{psbtrhdraftappendixeq3}
\end{equation}
of order $2n$ is self adjoint if and only if
\begin{itemize}
\item[(i)] the  even   coefficients $ r_0(x), r_2(x),\ldots,r_{2n}(x) $ are  continuous functions of $x$ and
\item[(ii)] the odd  coefficients $ r_1(x), r_3(x),\ldots,r_{2n-1}(x) $ satisfy the recurrence relation
\begin{small}
\begin{eqnarray}
r_{2k+1}=\frac{1}{2}\left\{\binom{2n}{2n-2k-1}r_{0}^{(2k+1)}-\binom{2n-1}{2n-2k-1}r_{1}^{(2k)}+\binom{2n-2}{2n-2K-1}r_{2}^{(2k-1)}-\cdots\right. \nonumber \\ \left. \cdots-\binom{2n-2k+1}{2n-2k-1}r_{2k-1}''+\binom{2n-2k-1}{2n-2k-1}r_{2k}'\right\}, \hspace{2cm} \label{psbtrhdraftappendixeq4} 
\end{eqnarray}
\end{small}
\end{itemize}
with initial condition $r_1=n\, r_0'$ and $k \geq 1$. A solution of the recurrence relation \eqref{psbtrhdraftappendixeq4} can be written as 
\begin{small}
\begin{eqnarray}
r_{2k+1}=\left\{\frac{1}{2} \binom{2n}{2k+1}+\sum_{l=2}^{k+1}\frac{(-1)^{l-1}}{2^l}\sum_{(k_1k_2\ldots k_l)\in I_{l}^{2k+1}}\binom{2n}{k_1}\binom{2n-k_1}{k_2}\cdots \binom{2n-\sum_{i=1}^{l-1}k_i}{k_l}\right\}r_{0}^{(2k+1)} \nonumber \\
+\left\{\frac{1}{2} \binom{2n-2}{2k-1}+\sum_{l=2}^{k}\frac{(-1)^{l-1}}{2^l}\sum_{(k_1k_2\ldots k_l)\in I_{l}^{2k-1}}\binom{2n-2}{k_1}\binom{2n-2-k_1}{k_2}\cdots \binom{2n-2-\sum_{i=1}^{l-1}k_i}{k_l}\right\}r_{2}^{(2k-1)}  \nonumber \\
+\left\{\frac{1}{2} \binom{2n-4}{2k-3} +\sum_{l=2}^{k-1}\frac{(-1)^{l-1}}{2^l}\sum_{(k_1k_2\ldots k_l)\in I_{l}^{2k-3}}\binom{2n-4}{k_1}\binom{2n-4-k_1}{k_2}\cdots \binom{2n-4-\sum_{i=1}^{l-1}k_i}{k_l}\right\}r_{4}^{(2k-3)} \nonumber \\
+\cdots+\left\{\frac{1}{2}\binom{2n-2k+2}{3}-\frac{1}{2^2}\binom{2n-2k+2}{1}\binom{2n-2k+1}{2}\right\}r_{2k-2}'''+\frac{1}{2}\binom{2n-2k}{1}r_{2k}', \hspace{2cm}
\label{psbtrhdraftappendixeq5}
\end{eqnarray}
\end{small}
where, for a positive odd integer $m$, the index set $I_{l}^{m}$ consists of the elements  $(k_1,k_2,\ldots,k_l)$ of length $l \geq 2$ defined by 
\begin{eqnarray}
(k_1k_2\cdots k_l) \in I_{l}^{m} \Leftrightarrow \begin{cases} k_1 \mathrm{\ is \  \ odd \  positive \ integer } \ < m, \\
k_1+k_2+k_3+\cdots+k_l=m, \\
k_2, k_3,\ldots ,k_l \mathrm{ \ all \ are \ even \ positive \ integers.}
\end{cases}
\label{psbtrhdraftappendixeq6}
\end{eqnarray}
Thus in  analogy with Eq.~\eqref{psbtrhdrafteq24}, the Lagrangian for the general equation \eqref{psbtrhdraftappendixeq3} is given by  
\begin{equation}
L(x,y,y',y'',\ldots,y^{2n})=y(x) \sum_{i=0}^{2n}r_i(x)\frac{d^{2n-i} y}{dx^{2n-i}}.
\label{psbtrhdraftappendixeq7}
\end{equation}  
The  expression  in  \eqref{psbtrhdraftappendixeq7} will  represent the Lagrangian function for the odd-order Riccati equations in \eqref{psbtrhdraftappendixeq1} and/or \eqref{psbtrhdraftappendixeq2} if $r_0(x)=1$ and $r_j(x)=\alpha_{2n-j}(x)$, $j=1,2,\ldots,2n-1$. Making use of Eq.~\eqref{psbtrhdrafteq17}  in Eq.~\eqref{psbtrhdraftappendixeq7} we finally obtain
\begin{small}
\begin{equation}
L(\cdot)=a^2 \, e^{2 \int \omega(x)\mathrm{d}x}\left(\vartheta^{2n-1}\omega+\alpha_{2n-1}(x)\vartheta^{2n-2}\omega+\alpha_{2n-2}(x)\vartheta^{2n-3}\omega\cdots+\alpha_{2}(x)\vartheta\omega+\alpha_1(x) \omega+\alpha_0(x)\right)
\label{psbtrhdraftappendixeq8}
\end{equation}
\end{small}
for the  $(2n-1)$th order Riccati equation.

\end{document}